\documentclass[showpacs,amsmath,amssymb,amsfonts,showkeys]{revtex4}

\usepackage{enumerate}
\usepackage{graphicx}% Include figure files
\usepackage{dcolumn}% Align table columns on decimal point
\usepackage[usenames]{color}
\usepackage{ulem} %\sout{strike out}
\providecommand{\abs}[1]{\left| #1 \right|} %

\def\x{\mathbf{x}}
\def\p{\mathbf{p}}
\def\X{\mathbf{X}}
\def\P{\mathbf{P}}

\begin{document}
%\draft
\date{\today}
\title{Do Tsallis distributions really originate from the finite baths?}

\author{G. Baris Bagci}
\affiliation{ Department of Physics, Faculty of Science, Ege
University, 35100 Izmir, Turkey}

\author{Thomas Oikonomou}\thanks{Corresponding Author}
\email{thoik@physics.uoc.gr}

\affiliation{Department of Physics, University of Crete, 71003
Heraklion, (Hellas) Greece}
\affiliation{Department of Physics, Faculty of Science, Ege
University, 35100 Izmir, Turkey}

%\pagenumbering{arabic}

\begin{abstract}
It is often stated that heat baths with finite degrees of freedom
i.e., finite baths, are sources of Tsallis distributions for the
classical Hamiltonian systems. By using well-known fundamental
statistical mechanical expressions, we rigorously show that
Tsallis distributions with fat tails are possible \textit{only}
for finite baths with constant negative heat capacity while
constant positive heat capacity finite baths yield decays with
sharp cut-off with no fat tails. However, the correspondence
between Tsallis distributions and finite baths holds at the
expense of violating equipartition theorem for finite classical
systems at equilibrium. Finally, we comment on the implications of
the finite bath for the recent attempts towards a $q$-generalized
central limit theorem.

\end{abstract}

%\narrowtext
\newpage \setcounter{page}{1}

\maketitle

\section{introduction}

Recently, a considerably great deal of effort has been put into
the construction of a non-extensive thermostatistics based on the
Tsallis entropies $S_{q}(p)=\frac{\int d\Gamma
p^{q}(\Gamma)-1}{1-q}$ where $\Gamma$ and $p(\Gamma)$ denote the
phase space variables and the probability distribution,
respectively \cite{Tsallis1}. The Tsallis $q$-distributions are
obtained from the maximization of the Tsallis entropies either by
the internal energy $E$ calculated from $E=\int d\Gamma p\left(
\Gamma \right) H$ or $E=\frac{\int d\Gamma p^{q}\left( \Gamma
\right) H}{\int d\Gamma p^{q}\left( \Gamma \right) }$ where $H$ is
the Hamiltonian. The former ones are called as ordinary Tsallis
distributions and of the form $1/\exp_q (\gamma_q H)$ ($\gamma_q$
being a positive constant) apart from normalization, where
$\exp_q(x)=\big[1+(1-q)x\big]_+^{\frac{1}{1-q}}$ with
$[x]_+=\max\{0,x\}$. The latter ones are escort Tsallis
distributions of the form $\exp_q (- \gamma_q H)$ omitting the
normalization constant.

Despite these efforts, however, the true origin of Tsallis
distributions in statistical mechanics is elusive. On the other
hand, there seems to be an agreement on an important and very
intuitive statistical mechanical source of Tsallis distributions,
namely, heat baths with finite number of degrees of freedom,
simply called finite baths \cite{Tsallis1, Plastino1, Almeida1,
Almeida2, Campisi1, Campisi2, Biro}.

According to this view, a system coupled to a finite heat bath
attains an inverse power law distribution in the form of
$q$-exponential decays for any of the two branches of the
non-extensivity parameter $q$, $q>1$ and $q<1$. The heat capacity
of the bath can be found as $\frac{1}{1-q}$ or $\frac{1}{q-1}$,
depending on which Tsallis distributions are used i.e., ordinary
or escort. Since the range of admissible $q$ values can be both
above and below the value $q=1$, the heat bath can have both
positive and negative heat capacity values. Despite seeming
counterintuitive at first, the possibility of systems with
negative heat capacity was first pointed out by Lynden-Bell and
Wood \cite{Bell} , later investigated theoretically by Thirring
who showed that microcanonical ensembles can in fact have negative
specific heats \cite{Th}. Recently, negative heat capacity
expressions have also been found in one dimensional evaporation
models and  long-range quantum spin systems in optical lattices
treated as microcanonical ensembles \cite{Hilbert, Kastner}. The
experimental measurements of the negative microcanonical heat
capacity have been carried out with small clusters of sodium atoms
near the solid to liquid transition and liquid to gas transition
for the cluster of hydrogen ions \cite{Schmidt, Gobet}.

As expected, when $q=1$, $q$-exponential decays become ordinary
exponential distributions and simultaneously the heat bath attains
infinite heat capacity \cite{Almeida1, Almeida2}.

In order to obtain the aforementioned results, one assumes that a
subsystem (simply called system from now on) embedded in a finite
heat bath interacts weakly, and together forms the total system.
Then, the total system can be treated microcanonically. In
treating subsystems in contact with an infinite reservoir, one can
either choose the phase space surface $\Omega$ or volume $\Phi$ as
the appropriate measure without loss of generality, since these
two yield the same results in the thermodynamic limit
\cite{Khinchin}. However, when dealing with systems of finite
degrees of freedom, one should consider two phase space measures,
separately, since these two measures might yield different
results. Therefore, a rigorous study of the finite baths must take
into account both ordinary/escort Tsallis distributions and phase
space volume/surface demarcation into account.

The paper is organized as follows: In the next section, we outline
the general microcanonical approach that will be used throughout
the manuscript. In Section III, we present the results by
considering all possibilities i.e., ordinary/escort distributions
as well as phase space volume/surface demarcations together with
the positivity or negativity of the total energy. Finally,
concluding remarks are presented.

\section{Microcanonical approach}

We begin by considering a system weakly coupled to another one
acting as a (finite) bath so that the total Hamiltonian is assumed
to be ergodic. Despite the presence of the interaction between the
system and bath, the total system i.e., system plus bath, is
assumed to be isolated. The Hamiltonian of the total system is
given as

%
%------------------------------------------------------------------------------
\begin{equation}\label{totalHam}
H_\text{tot}(\x,\p,\X,\P)=H_{\text{\tiny{S}}}(\x,\p)+H_{\text{\tiny{B}}}(\X,\P)
+h(\x,\X)\,,
\end{equation}
%------------------------------------------------------------------------------
%
where $H_\text{\tiny{S}}(\x, \p)$ and $H_\text{\tiny{B}}(\X,\P)$
denote the system and bath Hamiltonians, respectively, interacting
with one another through the interaction term $h(\x,\X)$. The
system and bath phase space coordinates are respectively given by
$\x=\{\x_1,\ldots,\x_{N_\text{\tiny{S}}}\}$,
$\p=\{\p_1,\ldots,\p_{N_\text{\tiny{S}}}\}$,
$\X=\{\X_1,\ldots,\X_{N_\text{\tiny{B}}}\}$
$\P=\{\P_1,\ldots,\P_{N_\text{\tiny{B}}}\}$ where
$\{\x_i,\p_i\}_{i=1}^{N_\text{\tiny{S}}},\{\X_i,\P_i\}_{i=1}^{N_\text{\tiny{B}}}
\in\mathbb{R}^{D}$, $D$ being the dimensionality of the space. The
system and bath Hamiltonians in particular read
%
%------------------------------------------------------------------------------
\begin{equation}\label{HamS}
H_\text{\tiny{S}}(\x,\p)=\sum_{i=1}^{N_\text{\tiny{S}}}\dfrac{\p_i^2}{2m}+
\mathcal{V}_\text{\tiny{S}}(\x)\,,\qquad\qquad
H_\text{\tiny{B}}(\X,\P)=\sum_{i=1}^{N_\text{\tiny{B}}}\dfrac{\P_i^2}{2M}+
\mathcal{V}_\text{\tiny{B}}(\X)\,,
\end{equation}
%------------------------------------------------------------------------------
%
where $\mathcal{V}_\text{\tiny{S}}(\x)$ and
$\mathcal{V}_\text{\tiny{B}}(\X)$ are the interactions within the
system and bath. From here on, the positivity of the system energy
$E_\text{\tiny{S}}$ is assumed. Moreover, the Boltzmann constant
is set equal to the dimensionless unity so that the temperature
has the same dimension as energy.

%The probability distribution $P(E_\text{\tiny{S}}>0)$ that the
%energy of the subsystem $S$ is equal to $E_\text{\tiny{S}}$ is
%given by
%
%------------------------------------------------------------------------------
%\begin{equation}\label{Prob1a}
%P(E_\text{\tiny{S}})=\dfrac{I_\text{\tiny{S}}(E_\text{\tiny{S}})
%I_\text{\tiny{B}}(\abs{E_\text{\tiny{B}}})}{I_\text{tot}(\abs{E_\text{tot}})}
%=\dfrac{I_\text{\tiny{S}}(E_\text{\tiny{S}})
%I_\text{\tiny{B}}(\abs{E_\text{tot}-E_\text{\tiny{S}}})}{I_\text{tot}(\abs{E_\text{tot}})}\,,
%\end{equation}
%------------------------------------------------------------------------------
%
%where $E_\text{tot}=E_\text{\tiny{S}}+E_\text{\tiny{B}}$ and
%$I:=\Omega,\Phi$ is either the phase space surface $\Omega$ or the
%volume $\Phi$ \cite{Khinchin}. They are related as
%
%------------------------------------------------------------------------------
%\begin{equation}
%\Omega(E)=\frac{\partial \Phi(E)}{\partial E}\,.
%\end{equation}
%------------------------------------------------------------------------------
%
%
The marginal probability distribution $ p(\x,\p)$ of finding the
system $S$ in a particular state with positive energy
$E_{\text{\tiny{S}}}$ reads \cite{Khinchin}
%
%------------------------------------------------------------------------------
\begin{equation}\label{MargProb0}
p( \x,\p) = \frac{\Omega _{\text{\tiny{B}}} \big(
E_\text{tot}-H_{\text{\tiny{S}}}( \x, \p) \big) }{\Omega
_\text{tot}( E_\text{tot} ) }=c\,\Omega _{\text{\tiny{B}}}\big(
E_\text{tot}-H_{\text{\tiny{S}}}( \x, \p ) \big)\,,
\end{equation}
%------------------------------------------------------------------------------
%
where $c$ is the normalization constant given by the inverse of
the density of states of the total system i.e.
%
%------------------------------------------------------------------------------
\begin{equation}\label{tot}
\Omega _\text{tot}\left( E_\text{tot}\right) =\int \text{d} \x\,
\text{d} \p\, \text{d} \X \,\text{d} \P \,\delta \big(
E_\text{tot}-H_\text{tot}\left( \x,\p, \X,\P \right)\big)\,.
\end{equation}
%------------------------------------------------------------------------------
%
$\delta$ denotes Dirac delta function. Similarly, the density of
states of the bath is given by
%
%------------------------------------------------------------------------------
\begin{equation}\label{bath}
\Omega _{\text{\tiny{B}}}\left( E_{\text{\tiny{B}}} \right) =\int
\text{d}\X\,\text{d}\P\, \delta \big( E_{\text{\tiny{B}}}
-H_{\text{\tiny{B}}}\left( \X,\P \right) \big),
\end{equation}
%------------------------------------------------------------------------------
%
where $E_{\text{\tiny{B}}}$ is the energy of the finite bath.
Before proceeding, we further assume that the finite bath has
constant heat capacity i.e. $C_{\text{\tiny{B}}}=
\frac{E_{\text{\tiny{B}}}}{T_{\text{\tiny{B}}}}$, implying in
particular that the heat bath is either composed of finite number
of non-interacting particles or particles interacting through
linear harmonic potential. However, the heat capacity of the
system, in full generality, is given by the expression
$C_{\text{\tiny{S}}}(T_\text{\tiny{S}})=\frac{\partial
E_{\text{\tiny{S}}}(T_\text{\tiny{S}})} {\partial
T_{\text{\tiny{S}}}}$, $T_{\text{\tiny{S}}}$ being the temperature
of the system. The system temperature becomes equal to that of the
bath only when the heat capacity of the bath is infinite.

The probability distribution $p(\x,\p)$ of the system in Eq.
\eqref{MargProb0}, due to the expression of the density of states
of the bath in Eq. \eqref{bath}, involves Dirac delta function
which is even, i.e., $\delta( x )=\delta ( - x )$. Therefore,
Dirac delta function enforces two distinct cases: either
$E_{\text{\tiny{tot}}}-H_{\text{\tiny{S}}}\geq0$ or
$E_{\text{\tiny{tot}}}-H_{\text{\tiny{S}}}<0$. Since the marginal
system distribution is finally obtained by identifying
$E_{\text{\tiny{B}}} = E_{\text{\tiny{tot}}}-H_{\text{\tiny{S}}}$,
one must consider both cases of the finite bath possessing
constant positive and negative energies, or due to the relation
$C_\text{\tiny{B}}=
\frac{E_\text{\tiny{B}}}{T_{\text{\tiny{B}}}}$, constant positive
or negative finite heat capacities. It is important to understand
the constraints imposed on the system probability distribution due
to these two possibilities: considering the case
$E_{\text{\tiny{tot}}}-H_{\text{\tiny{S}}}\geq0$, one can see that
the system energy at most can be equal to $E_{\text{\tiny{tot}}}$.
This in turn implies that the probability distribution of the
system in weak contact with a finite bath possessing positive heat
capacity must have a cut-off at
$H_\text{\tiny{S}}=E_{\text{\tiny{tot}}}$. This constraint
excludes the possibility of system distribution having fat tails.
Therefore, if fat power-law tails should ever emerge in this
context, this must be the case when the constant heat capacity of
the finite bath (or equivalently the heat bath energy
$E_\text{B}$) is negative, i.e.,
$E_{\text{\tiny{tot}}}-H_{\text{\tiny{S}}}<0$.

In order to proceed further, one must have an explicit expression
for the density of states of the bath composed of finite classical
particles. This expression can be shown to have the form $\Omega
_{\text{\tiny{B}}}\left( E_\text{\tiny{B}} \right) \sim
\abs{E_\text{\tiny{B}}}^{k}$ with exponent $k$ apart from some
multiplicative positive constant \cite{Almeida1, Almeida2,
Campisi3}. The absolute value is needed to ensure the positivity
of the density of states. Calculating the temperature of the
finite bath with $T_{\text{\tiny{B}}}^{-1} =
 \frac{\partial \ln
(\Omega _{\text{\tiny{B}}})}{\partial E_{\text{\tiny{B}}}}$ and
comparing it with the expression $C_{\text{\tiny{B}}}=
\frac{E_{\text{\tiny{B}}}}{T_{\text{\tiny{B}}}}$, we see that the
exponent $k$ is equal to the finite heat capacity of the bath
$C_{\text{\tiny{B}}}$ so that
%
%------------------------------------------------------------------------------
\begin{equation}\label{particularbath}
\Omega _{\text{\tiny{B}}}\left( E_{\text{\tiny{B}}} \right)
\sim
\abs{E_{\text{\tiny{B}}}}^{C_{\text{\tiny{B}}}}\,,
\end{equation}
%------------------------------------------------------------------------------
%
where the finite heat bath capacity can be positive (i.e.
$E_{\text{\tiny{tot}}}-H_{\text{\tiny{S}}}\geq0$) or negative
(i.e. $E_{\text{\tiny{tot}}}-H_\text{\tiny{S}}<0$).

At this point, we also remind the reader an important fact: one
should consider both the density of states $\Omega$ and the volume
of the phase space $\Phi$, since these two measures might yield
different results when dealing with systems of finite degrees of
freedom although they are equivalent in the thermodynamic limit
\cite{Dunkel}. The two measures are related to one another by

%------------------------------------------------------------------------------
\begin{equation}\label{twomeasures}
\Omega \left( E\right) =\frac{\partial\, \Phi( E) }{\partial E}\,.
\end{equation}
%------------------------------------------------------------------------------
%
The equation above yields $\Phi_{\text{\tiny{B}}} \left(
E_{\text{\tiny{B}}} \right) \sim
\abs{E_{\text{\tiny{B}}}}^{C_{\text{\tiny{B}}}+1} $ for a constant
heat capacity bath. Therefore, by using
$T_{\text{\tiny{B}}}^{-1}=\frac{\partial \ln (\Phi
_{\text{\tiny{B}}})}{\partial E_{\text{\tiny{B}}}}$ and comparing
with the expression $C_{\text{\tiny{B}}}=
\frac{E_{\text{\tiny{B}}}}{T_{\text{\tiny{B}}}}$, we obtain an
important relation

%------------------------------------------------------------------------------
\begin{equation}\label{relation}
C_{\text{\tiny{B}}}^{\Omega }+1=C_{\text{\tiny{B}}}^{\Phi },
\end{equation}
%------------------------------------------------------------------------------
%
which is valid as long as the heat capacity of the finite bath is
constant. The superscripts denote the quantities calculated
through the density of states
$\Omega\equiv\Omega_{\text{\tiny{B}}}$ or phase space volume
$\Phi\equiv\Phi_{\text{\tiny{B}}}$.

\section{Finite bath and Tsallis distributions}

Having outlined the general microcanonical approach to the
totality of the system plus bath, we now explore the constant
positive and negative heat capacity possibilities distinctly:

\vskip0.2cm

\paragraph{\hskip-0.3cm Case \hskip0.1cm
$E_{\text{\tiny{B}}}\geq0$}:
%
%\renewcommand{\theenumi}{\roman{enumi}}%
%\begin{enumerate}
%
%\item \textbf{Case} \textcolor{red}{$E_{\text{\tiny{B}}}\geq0$}:
%
This case i.e., $E_{\text{\tiny{tot}}}-H_\text{\tiny{S}}\geq0$,
corresponds to the marginal system distribution stemming from the
identification $E_\text{\tiny{B}} =
E_{\text{\tiny{tot}}}-H_\text{\tiny{S}}\geq0$. In other words, the
(finite) system is now coupled to a finite bath with positive
energy and therefore constant positive heat capacity
$C_\text{\tiny{B}}$. Therefore, the system energy can attain at
most the value $E_\text{\tiny{tot}}$, for which a necessary
cut-off condition has to be respected in the probability
distribution of the system. In fact, using Eqs. \eqref{MargProb0}
and \eqref{particularbath}, we obtain the marginal probability
distribution of the system
%
%------------------------------------------------------------------------------
\begin{equation}\label{positive1}
p( \x, \p ) = c \big( E_\text{\tiny{tot}}-H_\text{\tiny{S}}(\x,\p)
\big)^{C_\text{\tiny{B}}^{\Omega }}\,,
\end{equation}
%------------------------------------------------------------------------------
%
where $c$ is the normalization constant. This distribution can be
cast into the form of  escort $q$-exponential by identifying
$\alpha_q^{-1}:=(1-q)E_\text{\tiny{tot}}$ and
%
%------------------------------------------------------------------------------
\begin{equation}\label{positive2}
C_\text{\tiny{B}}^{\Omega }=\frac{1}{1-q}
\end{equation}
%------------------------------------------------------------------------------
%
so that
%
%------------------------------------------------------------------------------
\begin{equation}\label{positive3}
p( \x,\p) \sim \exp_q\big(-\alpha_q\, H_\text{\tiny{S}}( \x,\p)
\big)
\end{equation}
%------------------------------------------------------------------------------
%
apart from the normalization. Considering consistently the
conditions above together with $\alpha_q>0$, we obtain the
following range of validity for the non-extensivity index $q$
%
%--------------------------------------------------------------
\begin{equation}\label{positive4}
\left.%
\begin{array}{l}
    C_\text{\tiny{B}}^{\Omega }T^\Omega_\text{\tiny{B}}+E_\text{\tiny{S}}>0 \\
    E_\text{\tiny{S}}\;\,>0 \\
    C_\text{\tiny{B}}^{\Omega }T^\Omega_\text{\tiny{B}}\;>0 \\
    \alpha_q\;\;\; >0 \\
\end{array}%
\right\}\qquad\Longrightarrow\qquad
\left.%
\begin{array}{l}
    \frac{T^\Omega_\text{\tiny{B}}}{1-q}+E_\text{\tiny{S}}>0 \\
    E_\text{\tiny{S}}\;\,>0 \\
    \frac{T^\Omega_\text{\tiny{B}}}{1-q}\;>0 \\
    \frac{1}{1-q} >0 \\
\end{array}%
\right\}\qquad\Longrightarrow\qquad \boxed{q<1}\,,
\end{equation}
%--------------------------------------------------------------
%
with $T^\Omega_\text{\tiny{B}}>0$.
Accordingly, the distribution in Eq. \eqref{positive3} for $q<1$
represents a sharp decay with a cut-off at
$E_\text{\tiny{tot}}=H_\text{\tiny{S}}$ in the argument excluding
the possibility of fat tails.

On the other hand, in terms of the ordinary $q$-exponential
distributions, Eq. \eqref{positive1} can be rewritten as
%
%------------------------------------------------------------------------------
\begin{equation}\label{positive5}
p( \x, \p ) \sim \frac{1}{\exp_q\big(\epsilon_q\,
H_\text{\tiny{S}}( \x, \p)\big)}
\end{equation}
%------------------------------------------------------------------------------
%
with $\epsilon_q^{-1}:=(q-1)E_\text{\tiny{tot}}$ and
%
%------------------------------------------------------------------------------
\begin{equation}\label{positiveness1}
C_\text{\tiny{B}}^{\Omega }=\frac{1}{q-1}.
\end{equation}
%------------------------------------------------------------------------------
%
With the condition $\epsilon_q>0$ now, the following range of
validity for the index $q$ is found
%
%--------------------------------------------------------------
%--------------------------------------------------------------
\begin{equation}\label{positive6}
\left.%
\begin{array}{l}
    C_\text{\tiny{B}}^{\Omega }T^\Omega_\text{\tiny{B}}+E_\text{\tiny{S}}>0 \\
    E_\text{\tiny{S}}\;\,>0 \\
    C_\text{\tiny{B}}^{\Omega }T^\Omega_\text{\tiny{B}}\;>0 \\
    \epsilon_q\;\;\; >0 \\
\end{array}%
\right\}\qquad\Longrightarrow\qquad
\left.%
\begin{array}{l}
    \frac{T^\Omega_\text{\tiny{B}}}{q-1}+E_\text{\tiny{S}}>0 \\
    E_\text{\tiny{S}}\;\,>0 \\
    \frac{T^\Omega_\text{\tiny{B}}}{q-1}\;>0 \\
    \frac{1}{q-1} >0 \\
\end{array}%
\right\}\qquad\Longrightarrow\qquad \boxed{q>1}\,,
\end{equation}
%--------------------------------------------------------------
%
%
with $T^\Omega_\text{\tiny{B}}>0$.
The distribution in Eq. \eqref{positive6} for $q>1$ again
represents a sharp decay with a cut-off at
$H_\text{\tiny{S}}=E_\text{\tiny{tot}}$ in the argument excluding
the possibility of fat tails. These results show that the adoption
of ordinary or escort $q$-distributions does not change the form
of the probability distribution of the system. The choice between
the two aforementioned distributions is only related to the
intervals of $q$ values, since they are related to one another
through the relation $\left(2-q\right)$ (for more details on this
issue, see Refs. \cite{us1, us2}).
Note also that the same calculations above can be redone in terms
of $C_\text{\tiny{B}}^{\Phi }$ by using Eq. \eqref{relation}, but
it can be observed that the adoption of $C_\text{\tiny{B}}^{\Phi
}$ does not change the shape of the system distribution. Regarding
the related intervals in Eqs. \eqref{positive4} and
\eqref{positive6}, they change to $q<1 \wedge q>2$ and $q<0 \wedge
q>1$, respectively.

\paragraph{\hskip-0.3cm Case \hskip0.1cm
$E_{\text{\tiny{B}}}<0$}:
%
%\item \textbf{Case} \textcolor{red}{$E_\text{\tiny{B}}<0$}:
%
This case implies a finite bath with constant negative heat
capacity, since now $E_\text{\tiny{B}} =
E_\text{\tiny{tot}}-H_\text{\tiny{S}}<0$. To ensure the negativity
of the finite bath  energy, $E_\text{\tiny{tot}}$ must always be
negative so that the system distribution does not have a cut-off
now. Using again Eqs. \eqref{MargProb0} and
\eqref{particularbath}, the marginal probability distribution of
the system reads
%
%------------------------------------------------------------------------------
\begin{equation}\label{negative1}
p(\x, \p) = c\, \big( H_\text{\tiny{S}}(\x,\p)
-E_\text{\tiny{tot}} \big)^{C_\text{\tiny{B}}^{\Omega }}\,,
\end{equation}
%------------------------------------------------------------------------------
%
where $c$ is the normalization constant. This distribution can be
cast into the form of  escort $q$-exponential by identifying

%
%------------------------------------------------------------------------------
\begin{equation}\label{negative2}
C_\text{\tiny{B}}^{\Omega }=\frac{1}{1-q}
\end{equation}
%------------------------------------------------------------------------------
%
so that
%
%------------------------------------------------------------------------------
\begin{equation}\label{negative3}
p(\x, \p) \sim \exp_q\big(-\alpha_q\, H_\text{\tiny{S}}( \x, \p)
\big)
\end{equation}
%------------------------------------------------------------------------------
%
apart from the normalization. Considering consistently the
conditions above together with $\alpha_q>0$, we obtain the
following range of validity for the non-extensivity index $q$
%
%--------------------------------------------------------------
\begin{equation}\label{negative4}
\left.%
\begin{array}{l}
    C_\text{\tiny{B}}^{\Omega }T^\Omega_\text{\tiny{B}}+E_\text{\tiny{S}}<0 \\
    E_\text{\tiny{S}}\;\,>0 \\
    C_\text{\tiny{B}}^{\Omega }T^\Omega_\text{\tiny{B}}\;<0 \\
    \alpha_q\;\;\; >0 \\
\end{array}%
\right\}\qquad\Longrightarrow\qquad
\left.%
\begin{array}{l}
    \frac{T^\Omega_\text{\tiny{B}}}{1-q}+E_\text{\tiny{S}}<0 \\
    E_\text{\tiny{S}}\;\,>0 \\
    \frac{T^\Omega_\text{\tiny{B}}}{1-q}\;<0 \\
    \frac{1}{1-q} <0 \\
\end{array}%
\right\}\qquad\Longrightarrow\qquad
\boxed{1<q<1+\Big(\frac{E_\text{\tiny{S}}}{T_\text{\tiny{B}}}\Big)^{-1}}
\,,
\end{equation}
%--------------------------------------------------------------
%
with $T^\Omega_\text{\tiny{B}}>0$. As can be seen the term
$\frac{E_\text{\tiny{S}}}{T_\text{\tiny{B}}}$ has the dimension of
a heat capacity.
Accordingly, the distribution in Eq.
\eqref{negative3} represents an inverse power-law decay with fat
tails.
This is the result if one agrees to obtain temperature and
constant heat capacity of the finite bath in terms of the density
of states $\Omega$. In terms of the phase space volume $\Phi$,
using Eq. \eqref{relation}, we obtain
$C^\Phi_\text{\tiny{B}}=\frac{2-q}{1-q}$ so that Eq.
\eqref{negative4} yields
$1<q<1+(\frac{E_\text{\tiny{S}}}{T_\text{\tiny{B}}}+1)^{-1}$.
This result again indicates that the system distribution is an
inverse power-law with fat tails although in the thermodynamic
limit ($E_\text{\tiny{S}}\rightarrow \infty$ so that $q$ is
confined to the unique value of unity implying
$C_\text{\tiny{B}}^{\Omega}\rightarrow - \infty$), these
$q$-decays are replaced by the usual exponential distribution.

It is worth noting that a particular case of the above general
result was also obtained by Lutsko and Boon \cite{Lutsko}
\textit{solely} by considering the integration over momenta
degrees of freedom so that $E_\text{\tiny{S}}/T_\text{\tiny{B}} =
DN_\text{\tiny{S}} / 2$ (compare Eq. \eqref{negative4} above to
the one below Eq. (10) in Ref. \cite{Lutsko}) where $D$ and
$N_\text{\tiny{S}}$ denote the dimensionality of the phase space
and the number of particles in the system, respectively. It is
indeed very remarkable that Lutsko and Boon obtained this
particular result just by checking the integrability conditions of
the concomitant Tsallis distributions \cite{Lutsko}. However, the
results of Lutsko and Boon include neither the finiteness of the
bath nor the necessity of its negative heat capacity for the
escort $q$-distribution with fat tails to emerge. Due to the
generality of the present calculations, one can also consider the
influence of including other degrees of freedom on the interval of
validity of the non-extensive parameter $q$: the more the degrees
of freedom associated with the heat capacity of the system are,
the more confined is the interval of the possible $q$ values. As
such, the thermodynamic limit, for any system, corresponds to the
unique value of the non-extensivity parameter $q$ i.e. unity,
corresponding to the ordinary canonical case.

On the other hand, in terms of the ordinary $q$-exponential
distributions, Eq. \eqref{negative3} can be rewritten as
%
%------------------------------------------------------------------------------
\begin{equation}\label{negative6}
p( \x, \p ) \sim \frac{1}{\exp_q\big(\epsilon_q\,
H_\text{\tiny{S}}( \x,\p)\big)}
\end{equation}
%------------------------------------------------------------------------------
%
with
%
%------------------------------------------------------------------------------
\begin{equation}\label{negativeness1}
C_\text{\tiny{B}}^{\Omega }=\frac{1}{q-1}.
\end{equation}
%------------------------------------------------------------------------------
%
With the condition $\epsilon_q>0$ now, the following range of
validity for the index $q$ is found
%
%--------------------------------------------------------------
%--------------------------------------------------------------
\begin{equation}\label{negative7}
\left.%
\begin{array}{l}
    C_\text{\tiny{B}}^{\Omega }T^\Omega_\text{\tiny{B}}+E_\text{\tiny{S}}<0 \\
    E_\text{\tiny{S}}\;\,>0 \\
    C_\text{\tiny{B}}^{\Omega }T^\Omega_\text{\tiny{B}}\;<0 \\
    \epsilon_q\;\;\; >0 \\
\end{array}%
\right\}\qquad\Longrightarrow\qquad
\left.%
\begin{array}{l}
    \frac{T^\Omega_\text{\tiny{B}}}{q-1}+E_\text{\tiny{S}}<0 \\
    E_\text{\tiny{S}}\;\,>0 \\
    \frac{T^\Omega_\text{\tiny{B}}}{q-1}\;<0 \\
    \frac{1}{q-1} <0 \\
\end{array}%
\right\}\qquad\Longrightarrow\qquad
\boxed{1-\Big(\frac{E_\text{\tiny{S}}}{T_\text{\tiny{B}}}\Big)^{-1}<q<1}\,.
\end{equation}
%--------------------------------------------------------------
%
The distribution in Eq. \eqref{negative6} represents an inverse
power-law decay with fat tails by adopting the density of states
$\Omega$. In terms of the phase space volume $\Phi$, using Eq.
\eqref{relation}, we have $C_\text{\tiny{B}}^{\Phi
}=\frac{q}{q-1}$ so that
$1-(\frac{E_\text{\tiny{S}}}{T_\text{\tiny{B}}}+1)^{-1}<q<1$.
As expected, when the thermodynamic limit is attained, i.e.,
$E_\text{\tiny{S}}\rightarrow \infty$, the value of $q$ assumes
only unity, resulting an usual exponential decay.

Finally, we note that all the results in this Section rely on one
main ingredient: the dependence of the finite heat capacity of the
bath on the non-extensivity parameter $q$ (check Eqs.
\eqref{positive2}, \eqref{positiveness1}, \eqref{negative2} and
\eqref{negativeness1} for example). In fact, this is the sole
source for the emergence of the parameter $q$. However, the bath
has constant (positive or negative) heat capacity which implies
that it is composed of either finite number of non-interacting
particles or particles coupled with linear harmonic interaction.
In accordance with the equipartition theorem then, the constant
heat capacity of the finite bath is only proportional to the
degrees of freedom composing the bath with no explicit dependence
on $q$. Considering further that the equipartition theorem is
intact even for non-extensive systems \cite{Biro2}, it is apparent
that the non-extensivity parameter $q$ is merely originating from
a substitution for the finite heat capacity of the bath as opposed
to a possible genuine non-extensivity in the bath. In short, one
indeed has inverse power-law distributions due to the finiteness
of the heat capacity of the bath. However, these inverse power-law
distributions are not Tsallis distributions.\\

\section{Conclusions}

There has been a general consensus so far on relating finite baths
to the Tsallis distributions. According to this view, the finite
baths can have both positive and negative heat capacities
depending on the use of ordinary and escort probability
distributions \cite{Plastino1, Almeida1, Almeida2}. It is also
held that one can have inverse power law distributions of Tsallis
form with fat tails for all ranges of the nonextensivity parameter
$q$.

In order to shed light on all these issues, we have rigorously
studied the probability distribution of the system through
microcanonical approach and shown that it stems from the interplay
between any arbitrary system and the constant heat capacity of the
bath. Only when the bath has finite and constant negative heat
capacity, the system attains an inverse power law distribution
with fat tails. The finite baths with positive constant heat
capacity lead to the system distributions with a well-determined
cut-off condition leaving no possibility for the emergence of fat
tails.

Whether one adopts the ordinary or escort Tsallis distribution is
found to be irrelevant, since the choice between the two does not
change the nature of the distribution, but only serves for the
same distribution to emerge in different intervals of the
nonextensivity parameter $q$. Consider for example that one has a
thermal bath with finite and constant positive heat capacity.
Having this information suffices to determine the shape of the
system probability distribution i.e. a sharp decay with a cut-off
having no fat tails. However, the adoption of the ordinary Tsallis
distribution for this particular case limits the exponent of the
Tsallis distribution to the interval $q>1$ (see Eq.
\eqref{positive6}) while the same physical case yields $q<1$ for
the escort distributions (see Eq. \eqref{positive4}). Therefore,
it is the feature of the finite bath which determines the shape of
the system distribution.

The most important question is finally to decide on whether the
Tsallis distributions are indeed to emerge from the coupling of
the physical system with a finite bath. The answer to this is no,
since the emergence of the Tsallis distributions in finite bath
scenario, be it ordinary or escort, requires the constant heat
capacity of the finite heat bath to be $q$-dependent (see Eqs.
\eqref{positive2}, \eqref{positiveness1}, \eqref{negative2} and
\eqref{negativeness1} for a complete check). However,
 the bath, although finite because of consisting of finite number of non-interacting particles, should have a heat capacity such as $D N_B /2$, $D$ and $N_B$ being as usual the dimension of the space and the number of the particles in the bath, respectively. Therefore, its heat capacity must not depend on the non-extensivity
parameter
 $q$. One might argue that the bath, being finite, might not be extensive so that its heat capacity can be $q$-dependent to account for the degree of non-extensivity. However, it can be rigorously shown that
  the equipartition
 theorem is intact despite the non-additivity of the Tsallis
 entropies so that even the heat capacity of a non-extensive classical Hamiltonian system
 must be independent of the non-extensivity parameter $q$ \cite{Biro2}.

 One might object to the above conclusion by stating that the
 interaction energy between the system and the bath has not been
 taken into account. However, if this is done, the probability distribution
 is found to be neither exponential nor $q$-exponential \cite{Tay} (see Eq. (18)
 therein). We also note that our results do not exclude the
 possibility of Tsallis distributions in non-ergodic systems such
 as composed of classical long-range interacting particles \cite{central1}, since the ergodicity of the total system is assumed in the present work.

Considered in the context of the recent attempts to
$q$-generalized central limit theorems, our results on finite
baths seem consistent. There could be no $q$-generalized central
limit theorems if one could obtain Tsallis distributions from
ordinary classical systems coupled to finite baths. In other
words, if the finite baths were the source of genuine Tsallis
distributions, one would expect them to emerge from the ordinary
law of large numbers as an intermediate distribution despite the
lack of correlation. This is apparently not the case
\cite{central1, central2}.

\end{document}